# On the Formation of Thin Ice Crystal Plates: A New Type of Morphological Instability in Diffusion-Limited Growth

Kenneth G. Libbrecht[1]

*Norman Bridge Laboratory of Physics, California Institute of Technology 264-33, Pasadena, CA 91125*

**Abstract.** We propose a new type of morphological instability in the diffusion-limited growth of faceted crystals from the vapor phase that can explain the formation of thin ice plates at temperatures near -15 C. The instability appears when the attachment kinetics for facet growth depends strongly on the morphology of the facet surface. In particular, we propose that the condensation coefficient for growth of ice prism facets increases dramatically when the width of the facet approaches atomic dimensions. This model reconciles several conflicting measurements of ice crystal growth, and makes additional predictions for future growth experiments. Other faceted crystalline materials may exhibit similar morphological instabilities that promote the diffusion-limited growth of thin plate-like or needle-like crystal structures.

## 1. Introduction

The morphology of ice crystals growing in the atmosphere exhibits a complex and puzzling dependence on temperature. For example, at water vapor supersaturation levels often found in clouds, it is well known that freely falling ice crystals typically grow into thin plate-like forms at -2 C, long slender columnar crystals at -5 C, very thin plates again at -15 C, and columns again around -30 C (for a review of ice crystal growth from the vapor see [1]). These observations imply that the growth rates of the basal and prism facets of ice change dramatically over a rather small range in temperature, driven largely by changes in attachment kinetics at the facet surfaces. Despite more than a half-century of study, we still do not understand the basic physical mechanisms that are responsible for the unusual temperature-dependent morphology of growing ice crystals [1,2].

The focus of this paper will be on ice crystal growth from the vapor at temperatures near -15 C, where the most extreme plate-like growth is observed. Figure 1 shows an example of several thin ice plates that were grown in ordinary air at -15 C. Although the edges of the plates are not well resolved using optical micrography, we have made interferometric measurements of similar crystals that indicate that these plates are often only a few microns in thickness. Since the larger specimens in Figure 1 are over $100 \mu$ m in size, we find aspect ratios as high as $L_a/L_c \approx 50$ for these crystals, where $L_a$ is the crystal size along the $a$-axis (roughly equal to the diameter of the plate), and $L_c$ is the size along the $c$-axis (equal to the plate thickness).

In modeling the growth of these plate-like crystals, we begin by writing the crystal

---

[1] Address correspondence to *kgl@caltech.edu*; URL: http://www.its.caltech.edu/~atomic/

growth velocity in the usual Hertz-Knudsen form

$$v = \alpha \frac{\Omega(p_{surf} - p_{sat})}{\sqrt{2\pi mkT}} \quad (1)$$

where $p_{surf}$ is the water vapor pressure at the growing surface, $p_{sat}$ is the water vapor pressure in equilibrium with ice, $m$ is the mass of a water molecule, and $\Omega$ is the molecular volume. This relation defines the condensation coefficient, $\alpha$, that parameterizes the attachment kinetics at the ice surface. In general, $\alpha$ will be different for the two principal facets, and for fast kinetics $\alpha \to 1$.

We will see in the next section how detailed modeling of water vapor diffusion implies that $\alpha_{prism}/\alpha_{basal} \geq L_a/L_c$ under conditions leading to the growth of ice crystal plates like those shown in Figure 1. In the most extreme cases, we estimate that we must have $\alpha_{prism}/\alpha_{basal} \approx 100$ to produce such thin plates.

The problem with this simple inference is that some other measurements of ice crystal growth yield values of $\alpha_{prism}/\alpha_{basal}$ that are much smaller. For example, recent measurements of $\alpha_{prism}$ and $\alpha_{basal}$ as a function of supersaturation and temperature under near-vacuum conditions (i.e. with little background gas present other than water vapor) found $\alpha_{prism}/\alpha_{basal} \approx 3$ at $T = -15$ C, roughly independent of supersaturation over a substantial range [3]. Earlier measurements under similar conditions show considerable scatter [4-7], but none of these measurements yielded $\alpha_{prism}/\alpha_{basal}$ ratios high enough to explain the growth of thin plates like those in Figure 1. There appears to be a general unresolved discordance between measurements of $\alpha$ made in near-vacuum conditions and measurements made in air or other inert gases [8-11]

In the remainder of this paper, we first present more experimental evidence for this problem, and we then propose a mechanism that can explain the apparent discrepancies between the different measurements. We suggest additional ice crystal experiments that can confirm the existence of this growth mechanism, and we examine the role the mechanism may play in the growth of thin plate-like or needle-like crystal structures in other materials.

## 2. The Effects of Vapor Diffusion

To better understand the growth of thin plates like those in Figure 1, we grew plates under controlled conditions and measured the plate diameters $L_a$ as a function of growth time. In particular, we grew plates in ordinary air at -15 C with a supersaturation of $\sigma_\infty \equiv (p_\infty - p_{sat})/p_{sat} = 0.05$. The apparatus and experimental details were described in [12], and Figure 2 shows the measured plate sizes as a function of growth time. Following the procedures in [12], the points in Figure 2 represent only the largest plates observed. The fit line, given by $L_a(t) = 90(t/\tau)^{0.75}$ $\mu$m, where $\tau = 100$ seconds, defines an approximate upper limit on the observed crystal sizes. Many of the smaller crystals were either malformed in some way or were not exposed to the maximum supersaturation in the chamber [12]. The upper limit line thus represents the growth of near-perfect, isolated crystals in air at $\sigma_\infty = 0.05$.

Note that the morphology of ice crystals growing in air at -15 C depends on supersaturation [1], and the data in Figure 2 represent very thin, fast-growing simple hexagonal plate crystals. At low supersaturation levels, say for $\sigma_\infty \leq 0.01$, crystals growing in ordinary air at -15 C typically take the form of nearly isometric hexagonal prisms [1]. At somewhat higher supersaturations the growth changes to thicker plates, and the plate aspect ratio $L_a/L_c$ increases with increasing supersaturation. At $\sigma_\infty \approx 0.05$ we observe thin flat hexagonal plates like those shown in Figure 1, with the growth behavior shown in Figure 2. For larger crystals, or at substantially higher supersaturation levels, the growth morphology typically shows a complex dendritic structure. However, even at higher supersaturations the growth of the basal facets remains slow, so dendritic crystals growing at -15 C are still plate-like in their overall shape.

In addition to the supersaturation dependence, note also that thin plate-like ice crystals like those shown in Figure 1 can only be grown at temperatures close to -15 C [1]. Crystals grown at temperatures a few degrees higher or lower are still plate-like, but the aspect ratios are smaller than they are at -15 C. At temperatures near -8 C and -24 C, the growth forms are nearly isometric even for supersaturations near the water supersaturation level [1,13]. Crystals grown near -2 C can also be in the form of thin plates, but again the typical aspect ratios are substantially smaller than at -15 C.

From the fit line in Figure 2, we can infer the size $L_a$ and the growth velocity $v_{prism}$ of the prism facets as a function of time. This experiment did not yield measurements of $L_c$ for these crystals, but we performed additional interferometric measurements of similar crystals to obtain crude thickness information. We estimate that the basal growth velocity of these crystals was approximately $v_{basal} \approx 0.03$ $\mu$m/sec over the times shown in Figure 2. We estimate this velocity is only accurate to about a factor of two, but this is sufficient for the discussion here.

The growth of freely falling ice crystals is limited by a combination of effects: water vapor diffusion through the background gas, attachment kinetics at the different ice facets, and to a smaller extent heat diffusion that carries away latent heat generated by solidification. Ignoring crystal heating for the moment, we can model water vapor diffusion and attachment kinetics by solving the diffusion equation around the growing crystal. The problem is difficult for complex crystal morphologies, but becomes substantially simpler if we consider only the growth of simple hexagonal plates [14], and simpler still if we use a cylindrically symmetric disc as an approximation for a hexagonal plate. For the latter case we treat the edge of the disc as if it were a continuous prism facet, with attachment kinetics appropriate for a flat low-index surface.

Because the growth rates from the vapor are slow, the diffusion equation may be approximated by Laplace's equation for the supersaturation, $\nabla^2 \sigma = 0$. To define the necessary boundary conditions we can specify $\sigma$ on a boundary, or we can specify the growth velocity $v$ of an ice surface (essentially specifying $d\sigma/dx$ at the surface, where $x$ is distance along the normal to the surface), or we can specify $\alpha$ (which gives a relation between $\sigma$ and $d\sigma/dx$). The resulting 2D diffusion equation for cylindrically symmetric growth can be solved by Green's function methods [15,16], or by commercial finite-element codes [17]. Crystal heating was also included in our models by changing the effective $\sigma_\infty$, based on an estimate of the temperature rise of a freely falling crystal [12].

We examined solutions to the diffusion equation using a variety of reasonable boundary conditions, and further examined how sensitive the solutions were to uncertainties in our plate growth data. From these investigations we were able to obtain fairly reliable estimates of $\alpha_{basal}$ and $\alpha_{prism}$ under conditions in which thin plates grow. Figure 3 shows a schematic representation of the contours of constant supersaturation around a growing ice plate for a typical case. Because the growth rate of the prism facets is high, the solution of the diffusion equation indicates that $\sigma$ is low near the prism facets, and substantially higher near the centers of the basal facets. We estimate that $\sigma_{basal,max}$ is typically at least twice as large as $\sigma_{prism,max}$, where these represent the maximum values of the supersaturation at the basal and prism facets. This implies $\alpha_{prism}/\alpha_{basal} \approx 2 v_{prism}/v_{basal} \approx 2 L_a/L_c$ for crystals in the range shown in Figure 2. Again the factor of two is only a rough approximation, since this factor depends on the precise crystal thickness, morphology, and growth velocities input to the model.

The results of these calculations suggest that, for these very thin plates, the ice crystal cross section typically looks something like that shown in Figure 3 (where the vertical scale and molecular steps have been exaggerated). Since the supersaturation around the crystal is highest at the centers of the basal facets, that is where basal growth is nucleated. Molecular steps then propagate out toward the edges of the basal facets. The step density on the basal facets is higher near the edge of the crystal, where $\sigma$ is lower, resulting in the stable growth of this facet.

In our calculations we could not obtain a sufficiently high $v_{prism}$ using reasonable plate models without insisting that $\alpha_{prism} \approx 1$. With a substantially smaller $\alpha_{prism}$ the calculations cannot reproduce the growth data shown in Figure 2.

We also observed in our growth experiments that such large thin plate-like crystals were close to switching to dendritic growth, which begins when the prism surfaces become curved; some curvature of the growing surfaces can be seen in Figure 1. This behavior is consistent with $\alpha_{prism}$ being close to unity when these crystals are growing. Uncertainties in the plate thickness data do not allow us to constrain $\alpha_{prism}$ with high precision, but nevertheless we believe the preponderance of evidence points strongly toward $\alpha_{prism}$ being close to unity for the growth of these thin plate-like crystals.

## 3. Does Air Change $\alpha$ ?

The basic problem we are left with from the diffusion modeling is that we must have $\alpha_{prism}/\alpha_{basal} \approx 100$ to explain the growth of freely falling plate-like crystals in air, while direct measurements of $\alpha_{prism}$ and $\alpha_{basal}$ under near-vacuum conditions give only $\alpha_{prism}/\alpha_{basal} \approx 3$ [3]. One way we might explain this discrepancy is if the presence of the background gas somehow directly altered the attachment kinetics, so that $\alpha_{prism}/\alpha_{basal}$ was much larger in air than in near-vacuum.

Some previous measurements have suggested that a background gas of air or nitrogen does substantially change the attachment kinetics of ice, such that ice crystal growth is significantly impeded [10,11]. It is our opinion that these data are not completely reliable, partly because the earlier measurements are known to suffer from systematic errors [3], and because the procedure for obtaining the condensation coefficients from growth measurements (i.e. by solving the diffusion equation, as described above) is quite sensitive to systematic errors in the data [3]. Furthermore, in our opinion the proposed mechanism of "site poisoning" [10,11] seems unlikely in this case, given the low solubility of nitrogen and other gases in ice and water.

We explored this issue further by collecting additional data on the growth rates of ice crystals in air and in near-vacuum conditions. If site poisoning is to produce a large $\alpha_{prism}/\alpha_{basal}$, then in keeping with [10,11] we might expect that $\alpha_{basal}$ is substantially smaller in the presence of air, as compared to near-vacuum conditions. We therefore measured $\alpha_{basal}$ for crystals in air at a pressure of one bar, following the experimental procedures described in [3]. We made these measurements using small crystals, with diameters of roughly 20 $\mu$m, so that the corrections from vapor diffusion are fairly small [3]. The results, shown in Figure 4, clearly indicate that $\alpha_{basal}$ is essentially unaffected by the presence of a background gas of air, at least for the range of $\sigma_{surface}$ shown in Figure 4. Thus we conclude that the large observed value of $\alpha_{prism}/\alpha_{basal}$ is not because of site poisoning from the background gas, since we see no significant change in $\alpha_{basal}$ in the presence of air.

It seems even less likely that interactions with background gas could directly increase $\alpha_{prism}$, and furthermore such a hypothesis would not be consistent with the earlier measurements [10,11].

To the above discussion we can add another important measurement of ice crystal growth in background gases. Gonda [9] measured the morphology of ice crystals freely falling in background gases of helium and argon, as a function of gas pressure at -15 C. The supersaturation was fixed by a cloud of supercooled water droplets suspended in the gas, and was thus roughly independent of background pressure. These measurements revealed that thick, nearly isometric plates grew at pressures of ~0.25 bar, while thinner plates grew at pressures near one bar [9].

These results are consistent with everything presented above, as they show small $\alpha_{prism}/\alpha_{basal}$ ratios at low background gas pressures, and large ratios at higher pressures. We note that this qualitative result holds in air, argon, and helium, which suggests that air is no less inert than the noble gases insofar as its effect on the ice attachment kinetics.

## 4. Structure-Dependent Attachment Kinetics

The above data suggest a fairly simple picture for ice crystal growth from the vapor at -15 C. First, when the background gas pressure is low, approximately below 0.25 bar, $\alpha_{prism}/\alpha_{basal}$ is relatively small, so that freely-falling crystals grow into thick-plate hexagonal prisms, at least when the supersaturation is in the vicinity of five percent or higher. This appears to be true for a number of inert gases, including air.

Second, when the background pressure is near one bar, thin plate-like crystals grow at -15 C. Not all growing plates are extremely large and thin, but the data indicate $\alpha_{prism}/\alpha_{basal} \approx 100$ in the most extreme cases. This again appears to hold for a number of inert gases, including air. Quantitative measurements, combined with models of vapor diffusion in the background gas, indicate that $\alpha_{prism}$ is near unity for the growth of thin plates at gas pressures near one bar, while $\alpha_{prism} \ll 1$ under near-vacuum conditions (and perhaps whenever the pressure is below 0.25 bar). We likewise infer that $\alpha_{basal} \approx 0.01$ independent of gas pressure. (These are only rough estimates, since the condensation coefficients depend fairly strongly on supersaturation [3].)

So how do we reconcile these observations? We have considered the possibility that $\alpha_{prism}$ is somehow dramatically increased by direct interactions with the background gas, but this seems unlikely. There is no direct evidence that such an effect exists, and we know of no mechanism that might produce it. This possibility seems especially remote when considering the inert nature of the background gases and the small range of pressures in question.

We also considered the possibility that $\alpha_{prism}$ only appears to be large because of anomalous surface diffusion of admolecules. If water molecules readily diffused from the basal to the prism facet during crystal growth, then such diffusion could mimic a large $\alpha_{prism}/\alpha_{basal}$, even if $\alpha_{prism}$ was intrinsically small [18]. There is ample evidence that similar surface diffusion effects are present in metallic systems, and can lead to the growth of thin needle-like structures [19,20]. However, we do not feel surface diffusion between facets is responsible for the thin ice plates being considered here, mainly because of the dependence on background gas pressure. We expect surface diffusion would be largely unaffected by an inert background gas, yet thin plate-like crystals have never been observed to grow under low-gas-pressure conditions.

We propose a new model to explain the ice growth measurements, that we call *structure-*

*dependent attachment kinetics*, in which $\alpha_{prism}$ depends strongly on the structure of the prism facet under consideration. In particular, we propose that $\alpha_{prism}$ increases dramatically, to values near unity, when the width of the prism facet approaches atomic dimensions.

In this model the measurements described in [3] yielded relatively small values for $\alpha_{prism}$ because the crystal facets being measured were large in extent. These measurements yielded the intrinsic value of $\alpha_{prism}$ for a large prism facet, which is small.

In the case of freely falling thin plates, however, the width of the prism facets is approximately $w \approx \sqrt{aR}$, where $a$ is the step height and $R$ is the effective radius of curvature of the edge of the growing plate. Taking $R \approx 0.5 \mu$m for the edge of a growing plate, this yields $w \approx 40a$ in the case of ice. We are proposing that facets with such small widths will not exhibit the intrinsic value for $\alpha_{prism}$, but will instead exhibit $\alpha_{prism} \approx 1$.

There are a number of possible microscopic mechanisms that might lead to this type of structure-dependent attachment kinetics. A leading contender, in our opinion, is structure-dependent surface melting. We would expect surface melting to appear at lower temperatures for nanometer-wide facets than for macroscopic facets, in much the same way bulk melting occurs at lower temperatures in nanoscale molecular clusters [21]. Another possible mechanisms is enhanced surface roughening, again facilitated by the reduced binding forces present on a nanoscale facet.

Unfortunately these microscopic mechanisms are not readily calculable, given the complexity of the ice surface. Nevertheless, in our opinion it seems plausible that there exists a microscopic mechanism that could bring about the proposed structure-dependent attachment kinetics. We are not championing any specific microscopic mechanism here, but rather simply proposing a purely phenomenological model.

## 5. Indirect Effects from Diffusion

To see how structure-dependent attachment kinetics can explain the observed growth of ice crystals as a function of background pressure at -15 C, consider the growth of thin ice plates in the context of solvability theory, which describes diffusion-limited solidification under a wide range of conditions.

Starting with the case of high background pressure, we assume the growth is in the form of thin plate-like crystals, and examine the stability of that morphology. For thin plates our assumption is that $\alpha_{prism}$ is near unity because the width of the prism facet is very narrow. Thus we will have $\alpha \approx 1$ all around the growing edge of the plate, because the crystal surface is strongly curved and therefore can be considered nearly atomically rough.

Under such conditions 2D solvability theory [22,23] yields that the cross-section of the crystal will be roughly parabolic, with some edge radius $R$, as is shown schematically in Figure 3. The theory also indicates that this shape will be essentially independent of time (given that a growing disk is only a rough approximation to the 2D theory), and for the case of growth from the vapor [24] the edge radius will be approximately given by

$$R \approx \frac{2D}{s_0} \sqrt{\frac{2\pi m}{kT}}$$

where $D$ is the diffusion constant and $s_0$ is the dimensionless solvability parameter [22,23]. We note that solvability theory in this simple incarnation does not strictly apply to faceted crystals, and we expect that the shape near the plate edge will be distorted because of strong basal faceting. Nevertheless, we have found in previous measurements that solvability theory does give reasonable solutions in the case of ice [24], so we will proceed knowing that the above relation for $R$ is only approximate.

Now consider what happens when the background pressure is reduced. Since $D \sim P^{-1}$, where $P$ is background pressure, we see that the edge radius $R$ increases for lower $P$. As the pressure is lowered, we expect $R$ will eventually increase to some critical value, $R_{crit}$, at which point $\alpha_{prism}$ will revert to its normal value. When this happens, we would expect a rather abrupt transition to a thick-plate growth form, with atomically large prism facets.

The diffusion-limited growth of faceted crystals is still an unsolved theoretical problem, so we cannot explore this transition in detail. However the transition may easily be abrupt enough to explain the observed dramatic morphological change in ice growth between background gas pressures of 0.25 and 1 bar [9].

The dependence of ice crystal growth morphologies on supersaturation is more difficult to explain with this model, since solvability theory indicates that to lowest order $R$ is independent of $\sigma$. Here we suspect that higher order effects again increase $R$ with decreasing $\sigma$, so that the growth would again make a transition to more isometric forms as $R$ passes through $R_{crit}$. We are not able to better understand this transition with our approximate theory, however.

This model suggests that there may be a region of parameter space where freely falling crystals can grow as *either* thin plates or more isometric prisms, depending on the initial conditions that governed their growth. Thin plates could grow stably as long as the edge remained thin, so that $\alpha_{prism}$ remained high. If a crystal started out as a slightly thicker plate, however, then $\alpha_{prism}$ would be relatively small, and a more isometric prism would result. We have often observed thin plates and much thinker plates growing at the same time under what appeared to be identical conditions, which lends support to the model. However, additional measurements will be needed to explore this possible bimodal morphological distribution in a quantitative fashion.

# 6. Discussion

The evidence from ice crystal growth data at -15 C suggests that the attachment kinetics at the ice surface, as parameterized by the condensation coefficients for the two principal facets, depends very strongly on background gas pressure. When $\sigma_\infty \approx 0.05$, the evidence suggests that $\alpha_{prism}/\alpha_{basal}$ increases by nearly two orders of magnitude when the background gas pressure changes from below 0.25 bar to 1 bar. This result appears to hold for a background of air as well as several noble gases. We cannot explain this behavior by direct interactions between the gas and the ice surface, nor can it be explained by anomalous surface diffusion of water admolecules.

Our proposed phenomenological model of structure-dependent attachment kinetics can explain why $\alpha_{prism}/\alpha_{basal}$ changes so dramatically with only a relatively small change in background gas pressure. As shown above, the background gas is able to change $\alpha_{prism}$ in an indirect way, via the morphology of the growing ice crystals. Solvability theory provides an approximate theoretical framework to estimate the effective edge radius of the growing ice plate, which appears to be the critical parameter in determining $\alpha_{prism}$. Additional measurements of ice crystal growth from the vapor under varying conditions will be able to check the model in a more quantitative fashion than is currently possible.

We propose that structure-dependent attachment kinetics might be a fairly general phenomenon, applicable to many other materials. It will be most important for the diffusion-limited growth of plate-like or needle-like structures from the vapor phase. The phenomenon would be recognizable by a sharp transition in the growth morphology as a function of background gas pressure, as we observe with ice growth. Observations of structure-dependent attachment kinetics in different materials should lead to a better understanding the microscopic mechanisms underlying this phenomenon.

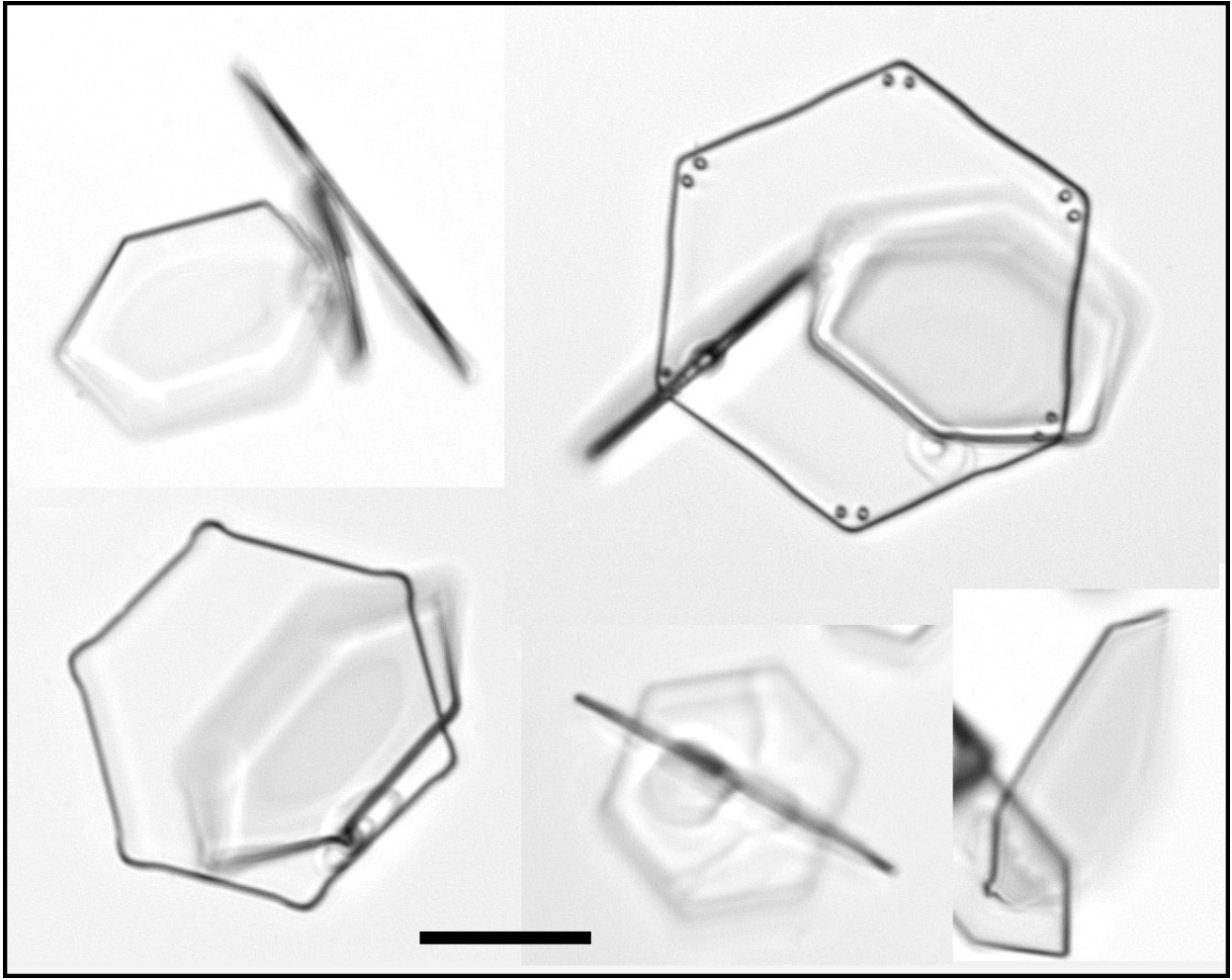

**Figure 1.** Photomicrographs of several plate-like ice crystals grown in ordinary air at a pressure of one bar and a temperature of -15 C. These crystals were grown in free-fall with growth times of roughly two minutes. The water vapor supersaturation level with respect to ice was approximately 0.05. The scale bar is 50 microns long.

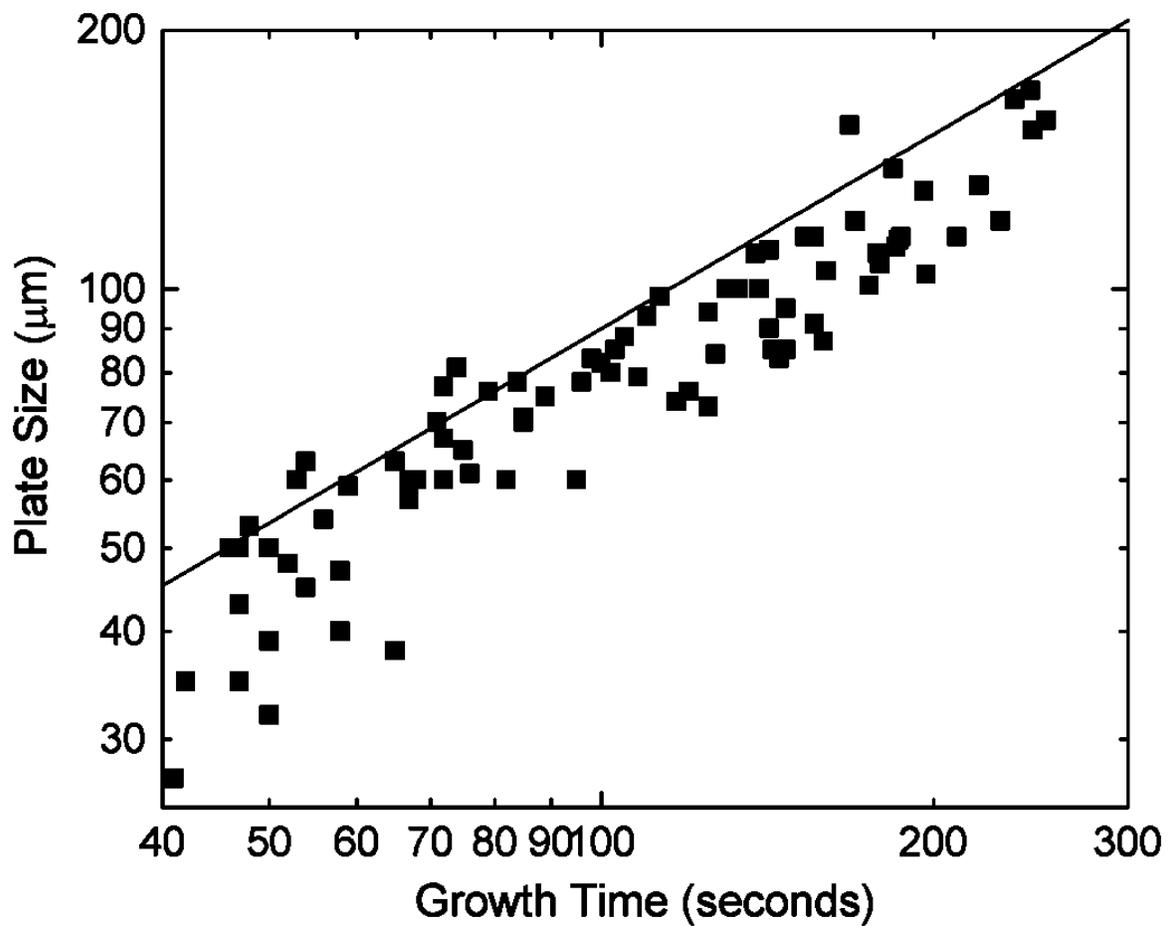

Figure 2. Plate size $L_a$ as a function of growth time, for freely falling ice crystals growing in air at a pressure of one bar, a temperature of -15 C, and a water vapor supersaturation with respect to ice of 0.05. The fit line represents an approximate upper limit to the crystal size [12].

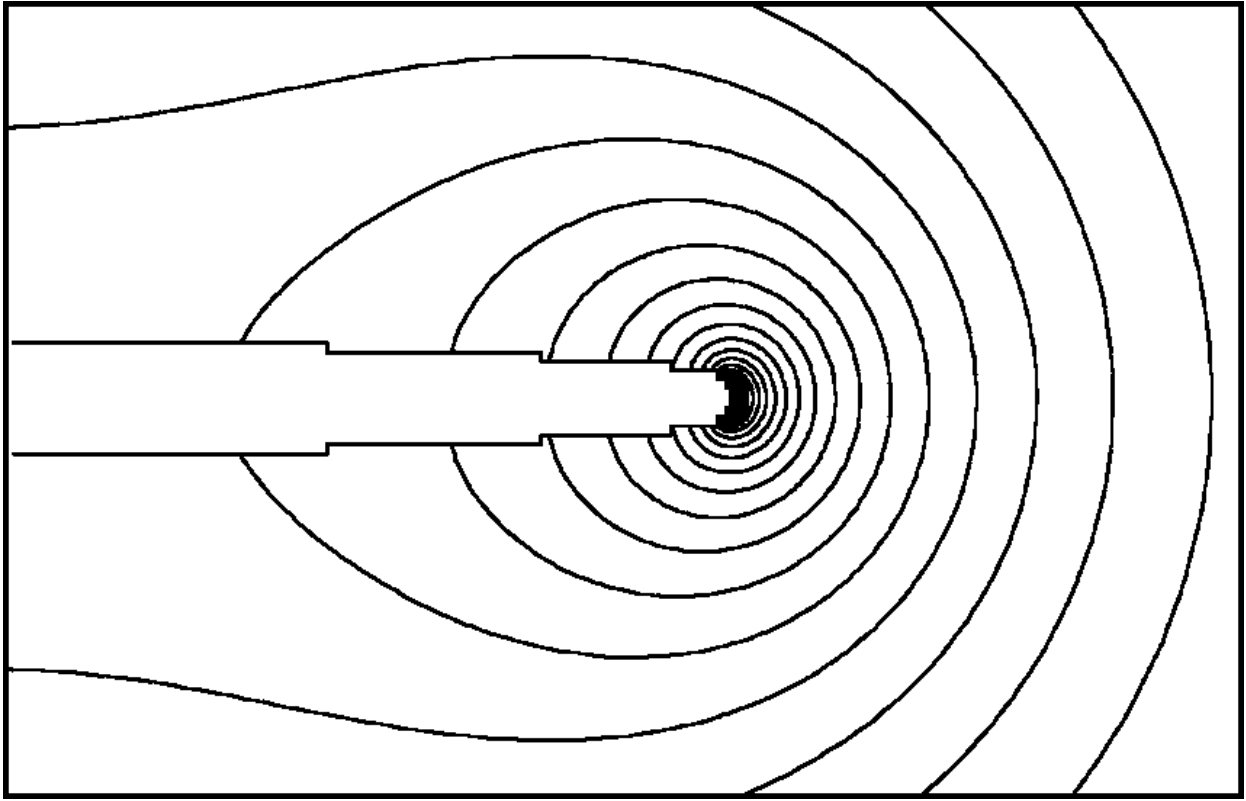

**Figure 3. Schematic representation of the profile of a growing ice plate, and contours of constant supersaturation around it; the symmetry axis is at the left edge of the figure. The vertical scale, and the molecular steps on the ice surface, have been exaggerated in this sketch. This supersaturation profile is determined by the large measured $v_{prism}/v_{basal}$ for thin plate-like crystals.**

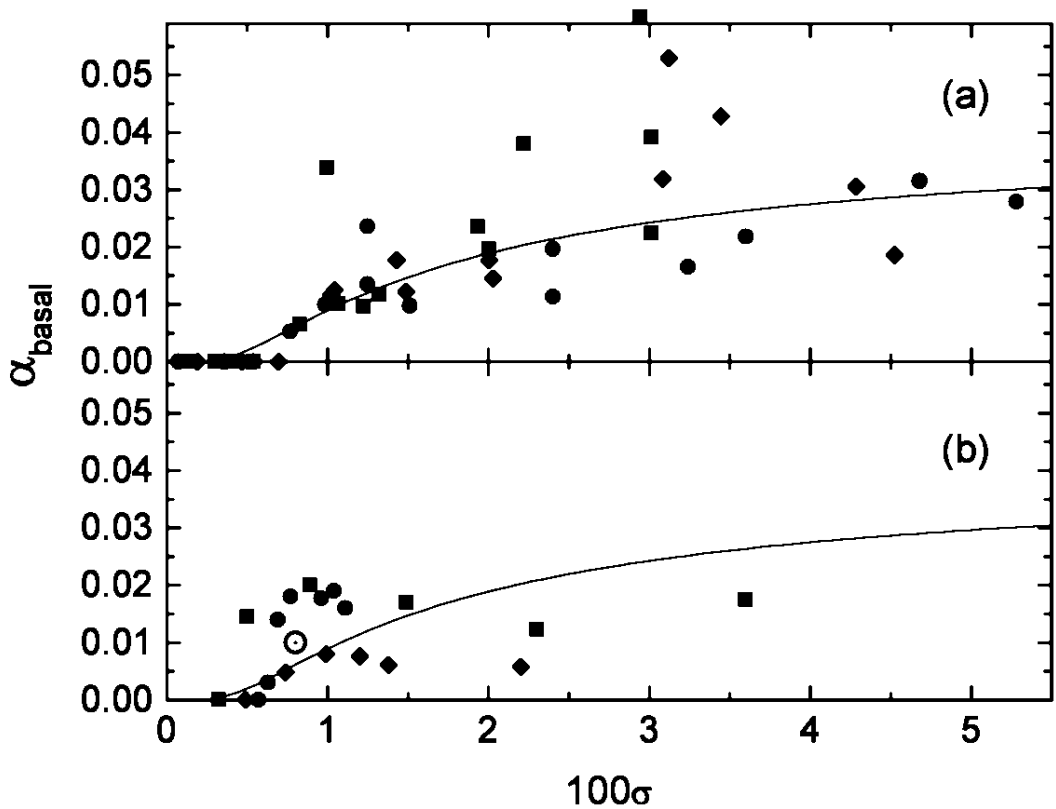

Figure 4. Measurements of the condensation coefficient for the basal facet as a function of water vapor supersaturation at the growing surface. The data and fit curve in (a) were described in [3], where growth velocity was measured in near-vacuum conditions. The data in (b) were acquired using similar techniques, but with the crystals growing in air at a pressure of one bar. The measured points have been corrected for the effects of vapor diffusion. The large open point in (b) was from an analysis of the plate growth data in Figure 2. The curve plotted in (b) is not a fit, but is identical to that in (a). Together these data indicate that the condensation coefficient shows little if any dependence on background gas pressure.